# An orthogonal-to-non-orthogonal multiplexing format converter

Authors: Zijian Li,[1,†] Chen Ding,[1,†] Zixian Wei,[2] Qiarong Xiao,[1] Ka-Suen Lee,[1] Chaoran Huang,[1] Changyuan Yu,[2] and Chester Shu[1,*]

[1] Department of Electronic Engineering and Centre for Advanced Research in Photonics, The Chinese University of Hong Kong, Shatin, New Territories, Hong Kong, China

[2] Department of Electrical and Electronic Engineering, The Hong Kong Polytechnic University, Kowloon, Hong Kong, China

[†] These authors contributed equally to this work

E-mail: ctshu@ee.cuhk.edu.hk

**Abstract text.**

Time-frequency orthogonality has been a foundational principle in the historical development of optical communications, whether in dense wavelength division multiplexing (WDM) within long-reach high-capacity coherent optical transmission or in time-frequency division multiple access within short-reach dense passive optical networks. Towards next-generation agile optical networks, jointly programmable orthogonal and non-orthogonal regulation offers flexible spectral allocation, ultra-dense packet distribution, and increased capacity. For bridging the fundamental differences of physical implementation, we propose and demonstrate a versatile orthogonal to non-orthogonal multiplexing format converter, with application to high-speed coherent optical transmission network enabled by a Talbot-based processor. The programmable Talbot-processed pumps coherently transfer and superpose optical signals of distinct wavelength channels onto a single channel through cross-phase modulation. We first demonstrate flexible conversion of two 80-Gbps WDM QPSK channels separated by 200–250 GHz into a non-orthogonal power-division multiplexing channel, while maintaining the high-quality encoded information in the digital domain. We then validate a digital-subcarrier-multiplexing dense access scenario in which eight 20-Gbps sub-channels are combined, converted, transmitted, and successfully decoded over a field-deployed fiber. The multiplexing format converter promises potential for applications in next-generation optical systems and networks with complex topologies and dense populations.

## 1. Introduction

Multiplexing is one of the foundational principles of modern information infrastructures [1-5]. By enforcing physical orthogonality between channels, for example, time- or frequency-domain multiplexing suppresses inter-channel interference and thereby enables multi-dimensional maneuverability. This principle underpins a wide range of bandwidth-hungry applications, including high-capacity optical communications, microwave photonics, optical computing, sensing and ranging, and quantum information [1-9]. The well-known wavelength-division multiplexing (WDM), which partitions the optical spectrum into orthogonal wavelength slots, has dramatically extended long-haul backbone transport capacity and intensified short-reach access network density [1, 10, 11]. Furthermore, optical frequency combs (OFCs) with mutually coherent, evenly spaced tones provide not only a precise spectral grid, but also a natural platform for comb-based transmission and all-optical processing. In this regard, OFC-based architectures are particularly appealing since they combine coherent carrier provisioning with a shared source for pump generation, regeneration, and signal processing. OFCs have further strengthened dense, low-latency, and cost-effective Internet scaling through precise and reconfigurable spectral grids [12-16].

Despite these advantages, the inherently channelized nature of WDM imposes a practical ceiling on spectral density and operational flexibility. Since each channel occupies a dedicated spectral bandwidth until decoded or processed, spectrum resources cannot be dynamically reused across users or traffic bursts, leading to rigid resource allocation, spectral fragmentation, and suboptimal utilization under dynamic, bursty traffic patterns [17]. This limitation is particularly acute in two critical real-world scenarios: coherent passive optical networks (PONs), where hundreds of distributed subscribers contend for a limited set of wavelengths at per-user data rates [18], and metro/data-center interconnects and edge access networks, where bursty, high-capacity traffic demands rapid reconfiguration and efficient multiuser aggregation [19].

Relaxing the constraint of strict orthogonality offers a compelling alternative to address these bottlenecks and enable next-generation reconfigurable optical networks [20-23]. Non-orthogonal multiplexing overlays multiple data streams onto a single carrier via distinct power levels, which are then recovered using digital successive interference cancellation (SIC) techniques. This ensemble enhances spectral efficiency (SE) and enables more flexible, dynamic resource allocation compared to orthogonal schemes [20, 24]. In current metro/access networks, a practical capability to convert orthogonal multiple-access (OMA) channels into non-orthogonal multiple-access (NOMA) channels would provide a seamless interface with existing WDM infrastructures and create a practical bridge that can exploit the strengths of both orthogonal and non-orthogonal multiplexing paradigms. Moreover, such compatible OMA-to-NOMA conversion can mitigate spectral fragmentation across heterogeneous optical networks, enable dynamic optical aggregation at the network edge, and support ultrahigh-bandwidth operation. It also maintains compatibility with existing OFC sources, fiber links, and coherent detection systems, making it suitable for scalable integration into current photonic architectures. All-optical signal processing is a particularly attractive route to realize orthogonal to non-orthogonal multiplexing format conversion, since the approach avoids electronic throughput bottlenecks and enables on-demand reconfiguration of aggregation granularity [25].

In this work, we propose and demonstrate an all-optical and reconfigurable orthogonal to non-orthogonal multiplexing format converter tailored for current coherent WDM networks. The conversion relies on Talbot-processed pulsed pumps and Kerr-induced cross-phase modulation (XPM) to coherently transfer signals from multiple frequency channels onto a single target wavelength, producing a power-superposed channel that remains compatible with standard coherent detection and digital signal processing (DSP)-based channel de-aggregation and multiuser recovery. By programming the pump spectral phase and thereby controlling the repetition-rate multiplication through different Talbot conditions, the processor can accommodate different channel spacings and aggregation granularities, enabling flexible, comb-aware operation across the WDM grid. The Talbot processor is therefore the key mechanism that enables precise and programmable repetition-rate multiplication for subsequent high-speed optical signal processing without relying on high-speed microwave electronics.

We validate our proposed converter with two proof-of-concept demonstrations, each tailored to specific real-world network scenarios. Firstly, targeting coherent PON uplink scenarios, we process two 80 Gbit/s WDM channels at the optical line terminal (OLT) after 20-km fiber transmission, aggregating them into a 160 Gbit/s non-orthogonal channel. Our approach cuts the number of coherent receiver ports required at the OLT, addressing the rigid resource allocation bottleneck in scalable PON deployments while ensuring recovery of all user information. Unlike conventional separate transmission and separate detection where each user requires an independent coherent receiver and dedicated DSP chain [1], our scheme performs channel aggregation directly in the optical domain before coherent reception. This shifts part of the multiuser resource management from the electrical layer to the physical optical layer, enabling receiver-port reduction, wavelength-slot reuse, and hardware consolidation at the aggregation node, where the complexity is localized and shared across multiple users, making the trade-off favorable for practical OLT-side deployment. Secondly, extending to dense edge access networks such as campus-scale access networks, we apply the orthogonal to non-orthogonal multiplexing format conversion to a denser scenario by converting two 80-Gbit/s digital-subcarrier-multiplexing (DSCM) channels, each consisting of four subcarriers, into a single nonorthogonal multiplexing DSCM channel, with validation over a 3-km field-deployed campus fiber link. Together, our demonstrations establish a practical pathway to boost capacity and flexibility in coherent optical networks by migrating part of multiuser resource management from electronics into the physical optical layer based on the concept of orthogonal-to-nonorthogonal multiplexing format conversion. Beyond communications, our concept also resonates with recent developments in intelligent optical systems and data-driven signal processing, where physical models are increasingly combined with adaptive algorithms to improve flexibility, robustness, and resource efficiency [26-28]. In addition, our proposed framework provides a programmable physical-layer substrate that can complement such approaches in optical computing and pattern recognition [21, 29, 30], where rethinking the constraints of channel orthogonality may enable novel functionalities.

## 2. Results

### 2.1. Principle of the orthogonal-to-non-orthogonal multiplexing format converter

Figure 1 illustrates the principle of the proposed all-optical orthogonal-to-non-orthogonal multiplexing format converter. A parametric OFC provides a set of mutually coherent, equally spaced optical carriers across a wide bandwidth, thereby enabling a high-channel-count WDM grid [31]. Within such comb-based WDM infrastructures, independently modulated wavelength channels can be processed directly in the analog optical domain and coherently merged onto a single channel in non-orthogonally power-superposed format (shown in Figure 1a). Two WDM signal channels with a frequency separation of $\Delta f_{ch}$, each carrying a modulated data sequence (for example, QPSK signal), expressed as $E_{ch1}\exp(j2\pi f_{ch1}t)$ and $E_{ch2}\exp(j2\pi f_{ch2}t)$, are routed into an all-optical processor (AOP). The AOP coherently transfers signals from both inputs onto a designated target frequency where two users are superimposed in the power domain. The composite optical waveform resembles a quasi-higher-order constellation, while the constituent user information remains separable by digital multiuser recovery. Reception proceeds with coherent detection and following DSP, where a single local oscillator suffices to downconvert the power-domain superimposed channel, and SIC algorithm is employed to recover the independent users in sequence. The stronger-power user is decoded first, its estimate is subtracted, and the weaker-power user is then recovered. In this way, the proposed converter brings the practical advantages of NOMA into the optical domain [20]. Two orthogonal WDM channels are recovered with a single coherent receiver, thereby reducing the required receiver-port count by half. At the same time, the two orthogonal wavelength slots are consolidated into a single non-orthogonal channel, freeing spectral resources for additional traffic or processing tasks and improving resource utilization and scalability at the aggregation node.

The AOP is implemented by programmable temporal Talbot processing of a pulsed pump followed by Kerr nonlinear interaction [32-35]. Figure 1(b) outlines the detailed operating mechanism. A pulsed pump derived from the same OFC can be written as $P_{pump}(t) = \sum_k a(t - k/f_r)$, where $f_r$ is the initial repetition rate, $a(t)$ represents the pulse envelope, and $k$ is an integer. The pump is routed through a programmable dispersive module whose group-delay dispersion is chosen to satisfy the temporal Talbot condition [36]:

$$\beta_2 z = \frac{1}{m}\frac{1}{2\pi {f_r}^2} \tag{1}$$

where $m$ is the repetition-rate multiplication factor. Subsequently, the pulse repetition rate is multiplied by an integer factor $m$, yielding a pulsed pump train with preserved pulse envelope, reduced timing jitter [37], and a new repetition rate of $mf_r$, expressed as $P_{pump_{new}}(t) \propto \sum_k a(t - \frac{k}{mf_r})$. The rate-multiplied pump is combined with the WDM signals and launched into a third-order ($\chi^3$) Kerr medium, inducing nonlinear phase shifts $\phi_{NL}$ to the channels, expressed as $\phi_{NL} = 2\gamma P_{pump} z$, with nonlinear parameter $\gamma$, effective interaction length $z$, and pump power $P_{pump}$. Under the small-signal approximation [38], the output field of channel $i$ acquires a phase modulation term that can be expanded in a Fourier series. The output field is expressed as:

$$E_{outi} \propto E_{chi} \sum_{n=-\infty}^{+\infty} c_n \exp(j2\pi(f_{chi} + nmf_r)t) \tag{2}$$

where the coefficients $c_n$ are determined by the pulse envelope and pump power, and quantify the power transferred to sidebands at frequency offsets $nmf_r$, which can be expressed as $c_n = mf_r \int_{-1/2mf_r}^{1/2mf_r} \exp(j2\gamma P_{pump} z) \exp(-jn2\pi mf_r t)\, dt$. Eq. (2) represents the frequency channel can be shifted by $nmf_r$ with a scaling coefficient of $c_n$. When the channel separation $\Delta f_c = |f_{ch1} - f_{ch2}|$ equals an integer multiple of the multiplied repetition rate (i.e., $\Delta f_c = q \cdot mf_r, q \in Z$), the XPM mapping aligns the distinct frequency channels to the same frequency bin, producing an all-optical signal aggregation in the power domain. By programming the multiplication factor $m$, the AOP can accommodate different channel spacings and signal formats, which makes the conversion reconfigurable across a wide range of practical WDM conditions. In addition, several AOP units may operate in parallel across a comb grid to convert multiple channel pairs simultaneously, enabling scalable aggregation. Our proposed architecture reclaims spectral resources by consolidating carriers, reduces hardware complexity at the receiver side, and preserves interoperability with existing comb sources, fiber deployments, and coherent receivers. These attributes render the proposed approach attractive for heterogeneous environments ranging from metro and data-center interconnects to access PONs.

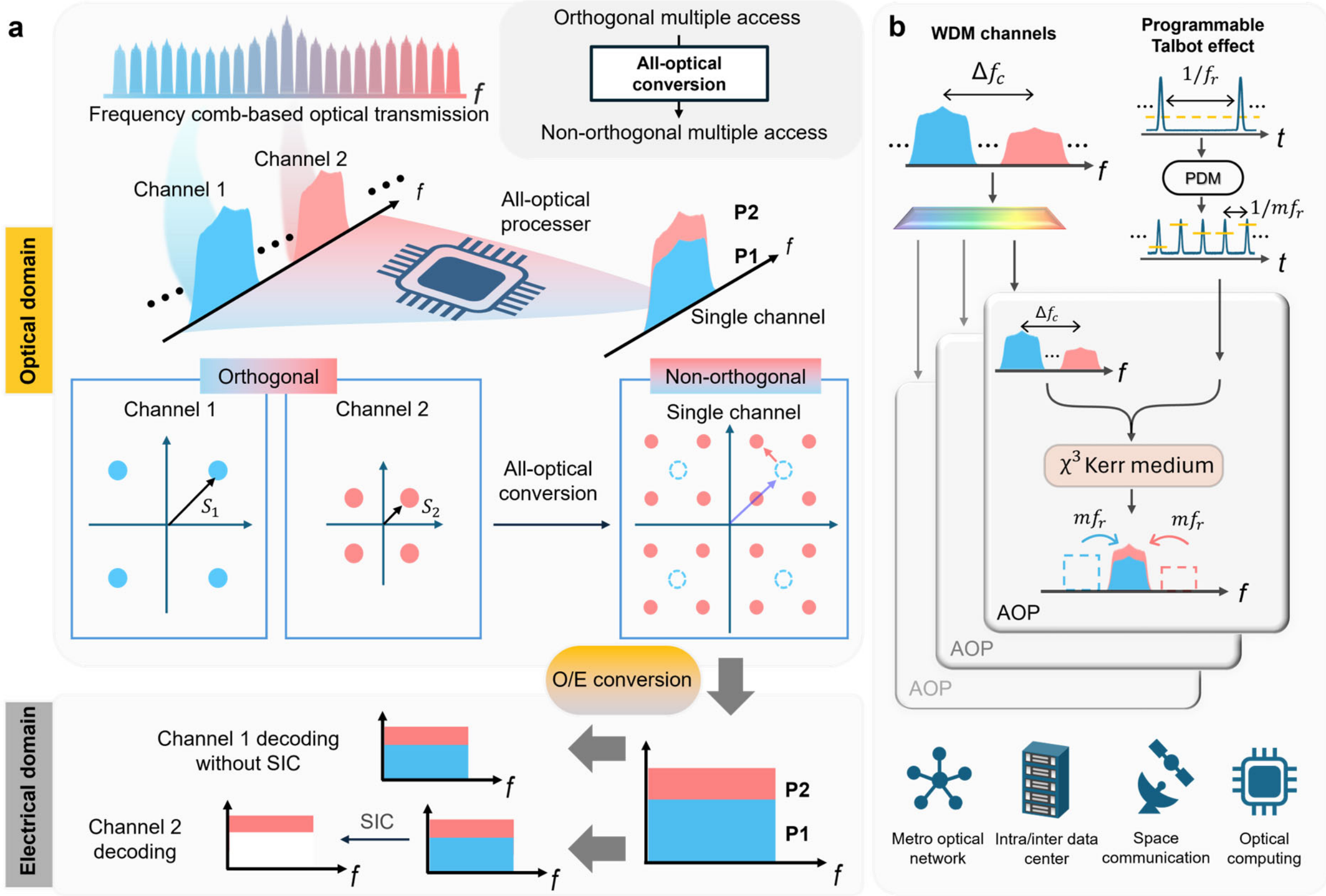

Figure 1 **a** Schematic overview. A parametric optical frequency comb (OFC) provides a dense grid of mutually coherent carriers for WDM. Two WDM channels (shown in blue and red), each carrying an independent high-speed signal are routed into an all-optical processor (AOP) that coherently transfers the signals onto a single target wavelength, producing a power-superposed, non-orthogonal channel. A coherent receiver downconverts the aggregated channel. Successive-interference-cancellation (SIC) is used to recover the higher-power user first while treating the remaining users as additional interference. The detected waveform is digitally reconstructed and subtracted from the composite received signal, thereby suppressing its interference contribution. The residual signal, dominated by the weaker user, is then demodulated using the same digital signal processing chain. **b** Implementation principle based on programmable temporal Talbot processing and Kerr-induced ($\chi^3$) cross-phase modulation (XPM). Two WDM carriers separated by $\Delta f_c$ are combined with a pulsed pump of base repetition rate $f_r$. The pump is routed through a programmable dispersive module (PDM) that satisfies various temporal Talbot conditions to multiply the repetition rate by an integer $m$. When $\Delta f_c$ is an integer multiple of $mf_r$, XPM in the Kerr medium coherently maps the carriers to the same spectral position so they aggregate in the power domain. The AOP is reconfigurable by programming $f_r$ and the Talbot multiplication factor $m$ to accommodate different channel spacings and signal formats.

It is necessary to distinguish our proposed multiplexing-domain conversion from conventional all-optical modulation format convertors (MFCs) [29, 39-48]. An all-optical MFC is primarily concerned with modulation-format transformation (e.g., mapping on–off keying (OOK) to QPSK or converting QPSK into higher-order constellations). In such schemes, optical fields are coherently combined to synthesize a new symbol alphabet, which typically requires stabilization of relative carrier phases to ensure deterministic constellation formation. In addition, the relative power ratio between the original modulation-format components must usually be carefully controlled to realize a target constellation, such as 16-QAM. Although the final waveform may also be viewed as a power superposition of the original fields, it is typically not treated as a well-defined non-orthogonal multiplexing format that supports independent recovery of the original channels via standard SIC-based processing.

By contrast, our proposed all-optical orthogonal-to-nonorthogonal multiplexing format conversion operates in the multiplexing domain rather than the modulation format domain.

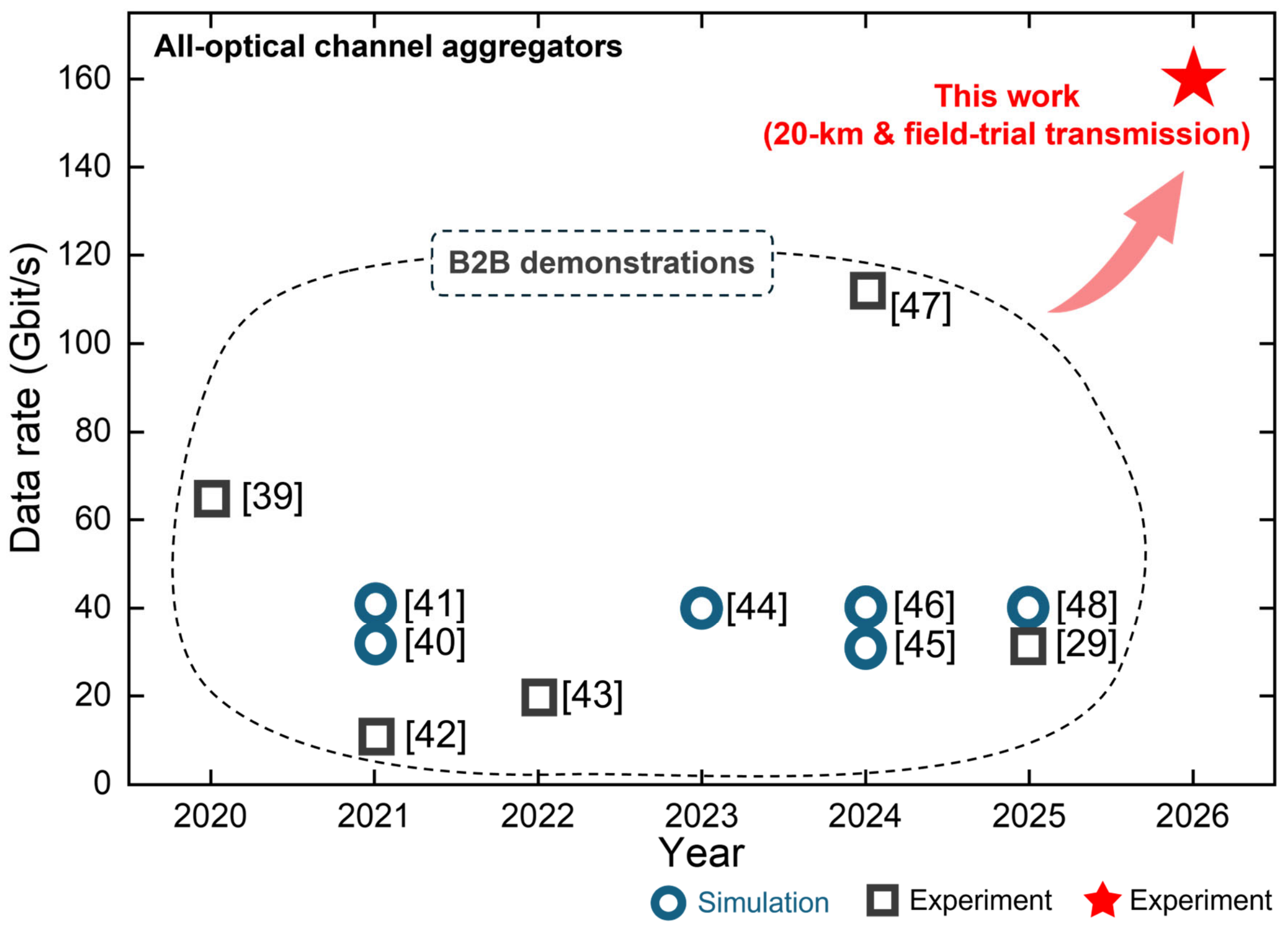

Figure 2 Comparison of processing data rate between our work and other all-optical channel aggregators. [29, 39-48]

Different wavelength channels are aggregated into a single power-domain-multiplexing channel while preserving separable channel information. The relative carrier phases between users are neither controlled nor required to be stable; instead, the relative carrier phases naturally appear as part of the multiuser interference. During further SIC algorithim, this interference contributes to the effective noise level when decoding the stronger-power user and is subsequently mitigated once the dominant signal is estimated and subtracted. Consequently, inter-carrier phase offsets do not impose stringent operational constraints and are fully accommodated within the digital domain.

From a system perspective, our proposed scheme therefore serves a different role from all-optical MFCs. Rather than synthesizing a new modulation format, it acts as an interface between orthogonal wavelength multiplexing and power-domain multiple access, enabling optical aggregation that remains compatible with comb-based WDM transmission infrastructures and standard coherent receivers. The system-level comparisions between MFCs and our proposed all-optical orthogonal-to-nonorthogonal multiplexing format conversion are summarized schematically in Fig. 2.

**2.2. Versatile orthogonal to nonorthogonal multiplexing format conversion**

We perform a proof-of-concept experiment to demonstrate our proposed orthogonal-to-non-orthogonal multiplexing format converter, which addresses the critical demand for scalable transmission in the multipoint-to-point (MP2P) uplink scenario of coherent PON architectures. Figure 3a depicts the experimental setup. A parametric OFC is generated from a narrow-linewidth (< 0.1 kHz) continuous-wave laser (NKT BASIK E15) cascaded with an intensity modulator and a phase modulator driven by a radio-frequency generator, followed by an erbium-doped fiber amplifier (EDFA) and a highly nonlinear fiber, producing a parametric frequency comb with an FSR of 25 GHz that served as the optical carrier source (see Supplement S1). In a practical coherent PON architecture, the optical frequency comb is generated at the OLT and distributed downstream to optical network units (ONUs), where individual carriers are modulated with user data and transmitted back upstream [49]. To emulate this scenario, we select two comb lines at 1555.35 nm (blue) and 1556.96 nm (red) using a programmable optical filter (POF), representing carriers assigned to two ONUs for WDM-based uplink transmission. Each carrier is independently modulated to produce an 80 Gbit/s QPSK signal using an IQ modulator. The two coherent WDM channels are then combined and transmitted over a 20-km standard single-mode fiber (SSMF), mimicking simultaneous uplink signals from multiple ONUs to the OLT.

The optical pulsed pump driving the multiplexing format conversion is derived from the same comb source via spectral slicing through the POF, centered at 1540 nm. The initial pump repetition rate is 25 GHz, as shown in Fig. 3b. Given the 200 GHz frequency separation between the two WDM channels, the pump repetition rate must be increased to 100 GHz to satisfy the condition for efficient multiplexing conversion. This is achieved using a programmable temporal Talbot processor (see Supplement S2), whose total dispersion is adjusted to meet the target Talbot condition for $m = 4$ (see Eq. 1). After temporal Talbot effect, the pump repetition rate is quadrupled from 25 GHz to 100 GHz, as experimentally verified in Fig. 3b. A comparison of the optical spectra before and after Talbot processing is provided in Supplementary S3. The processed pump, with 17 dBm power, and the two transmitted WDM-based QPSK signals are then jointly launched into a nonlinear Kerr medium located at the OLT. Through XPM, the pump induces coherent spectral translation that transferred the signals at both channels onto a common target wavelength at 1556.16 nm (see Fig. 3c, with full spectra shown in Supplementary S4). After nonlinear interaction, a dense wavelength-division multiplexer (DWDM) filter is used to isolate the converted signal channel centered at 1556.16 nm (Fig. 3c, bottom), corresponding to a power-domain superposition of the two original ONU signals. The converter reduces the required coherent receiver count from two to one at the OLT, corresponding to a 50% reduction in receiver-port count and a two-to-one consolidation of wavelength-slot usage. This converted non-orthogonal channel is received by a single coherent receiver, and the information of both original WDM channels is recovered via DSP incorporating SIC. The receiver-side DSP flow is summarized in Supplementary S2. Figure 3d shows the bit-error-rate (BER) performance as a function of the optical power allocation ratio between ONU-1 and ONU-2, with a total data rate of 160 Gbit/s and a fixed received optical power (ROP) of 7 dBm for the aggregated channel. The power allocation ratio is tuned using a variable optical attenuator (VOA) while considering the nonlinear conversion efficiency (see Supplement S5). It should be emphasized that the VOA is not introduced to artificially penalize one channel, but to realize the controlled power allocation required for power-domain multiplexing. In practical systems, such power imbalance naturally arises from differences in path loss, transmission distance, or, more generally, unequal channel conditions. In our proof-of-concept experiment, the VOA provides a controllable and repeatable means to emulate these conditions and to investigate the optimal power allocation ratio for stable channel recovery.The results reveal that as the ratio increases, ONU-1 (the higher-power user) benefits from improved signal-to-interference-plus-noise ratio (SINR), whereas ONU-2 experiences progressively degraded performance due to residual interference propagation. At the optimal ratio of 7.5 dB, both ONUs achieved BERs below the soft-decision forward error correction (SD-FEC) threshold of $2.2 \times 10^{-2}$, confirming effective multiuser coexistence within a single non-orthogonal channel.

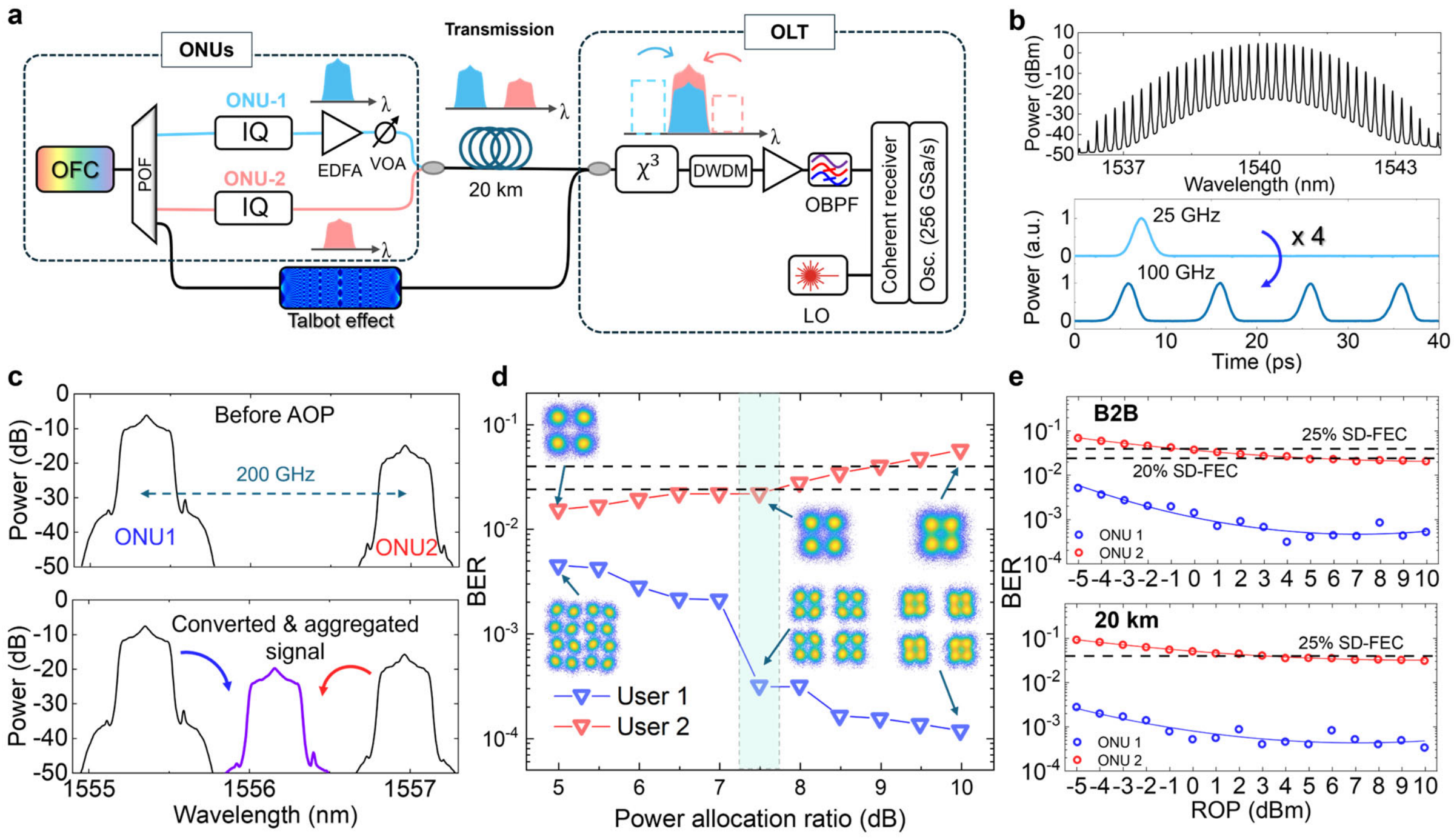

Figure 3 **a** Schematic illustration of the proof-of-concept system. OFC: optical frequency comb; IQ: In-phase and quadrature modulator; POF: programmable optical filter; EDFA: erbium-doped fiber amplifier; VOA: variable optical attenuator; DWDM: dense wavelength-division multiplexer; OBPF: optical bandpass filter; LO: local oscillator. **b** Talbot-processed optical pump derived from the same comb source. The repetition rate of the pulsed pump is multiplied from 25 GHz to 100 GHz through a programmable Talbot module that satisfies the Talbot condition as $m = 4$. **c** Optical spectra measured before the all-optical signal processing and at the output of the all-optical signal processor, showing the converted power-domain non-orthogonal signal centred at 1556.16 nm after cross-phase modulation (XPM). **d** Bit-error-rate (BER) of user 1 and user 2 versus optical power allocation ratio at a total data rate of 160 Gbit s$^{-1}$ and a received optical power (ROP) of 7 dBm. **e** BER versus ROP for a fixed 7.5 dB power allocation ratio.

To further evaluate transmission robustness, BER performance is analyzed as a function of ROP for the optimal power allocation ratio, as shown in Fig. 3e. As ROP increases from −5 dBm to 10 dBm with a step of 1 dBm, both ONUs exhibit improvement in BER for back-to-back (B2B) and 20 km transmission cases. For the B2B configuration, both users maintained BERs below $2.2 \times 10^{-2}$ when ROP was between 5 dBm and 10 dBm. After 20-km fiber transmission, the BERs remain below $4 \times 10^{-2}$ for ROPs between 4 dBm and 10 dBm, which is within 25% SD-FEC limits. The small performance gap between the B2B and fiber-transmission cases indicates that the proposed all-optical multiplexing format conversion maintains high signal fidelity and exhibits tolerance to transmission-induced impairments, including linear dispersion (which manifests as deterministic phase distortion and can be compensated by coherent DSP), as well as fiber-induced phase noise and nonlinear distortions [50].

To demonstrate the reconfigurability of our system, we reconfigure the AOP to accommodate a different WDM spacing and evaluate the resulting conversion fidelity and system performance. In this experiment, the two input carriers are separated by 250 GHz (ONU-1 at 1555.16 nm and ONU-2 at 1557.17 nm). The pulsed pump is therefore processed under a Talbot condition with repetition-rate multiplication factor $m = 5$, where the effective dispersion required for the Talbot condition is synthesized by additional line-by-line spectral shaping using the POF [35]. Figure 4a presents the optical spectra before and after conversion for a fixed power allocation ratio (PR = 7.5 dB), showing that the original signals modulated on the two carriers are coherently transferred to the target wavelength at 1556.16 nm. Temporal traces in Fig. 4b compare the pump waveform prior to temporal Talbot effect and after pulse-rate multiplication with $m = 5$. The Talbot processor multiplies the pulse rate while preserving the overall envelope of the comb spectrum. The power of the processed pump power is amplified to a slightly higher level at 18 dBm. The output spectrum of the pulsed pumps after the AOP is shown at the bottom of Fig. 4b. Figure 4c summarizes BER versus ROP for $m = 5$ under both B2B and 20-km transmission conditions. The BER trends mirror those obtained for the $m = 4$ case, exhibiting monotonic improvement with increasing ROP in which both users meet the SD-FEC threshold. The strong consistency across different WDM channel spacings confirms that the proposed scheme can flexibly adapt to varying channel conditions without introducing significant performance penalties.

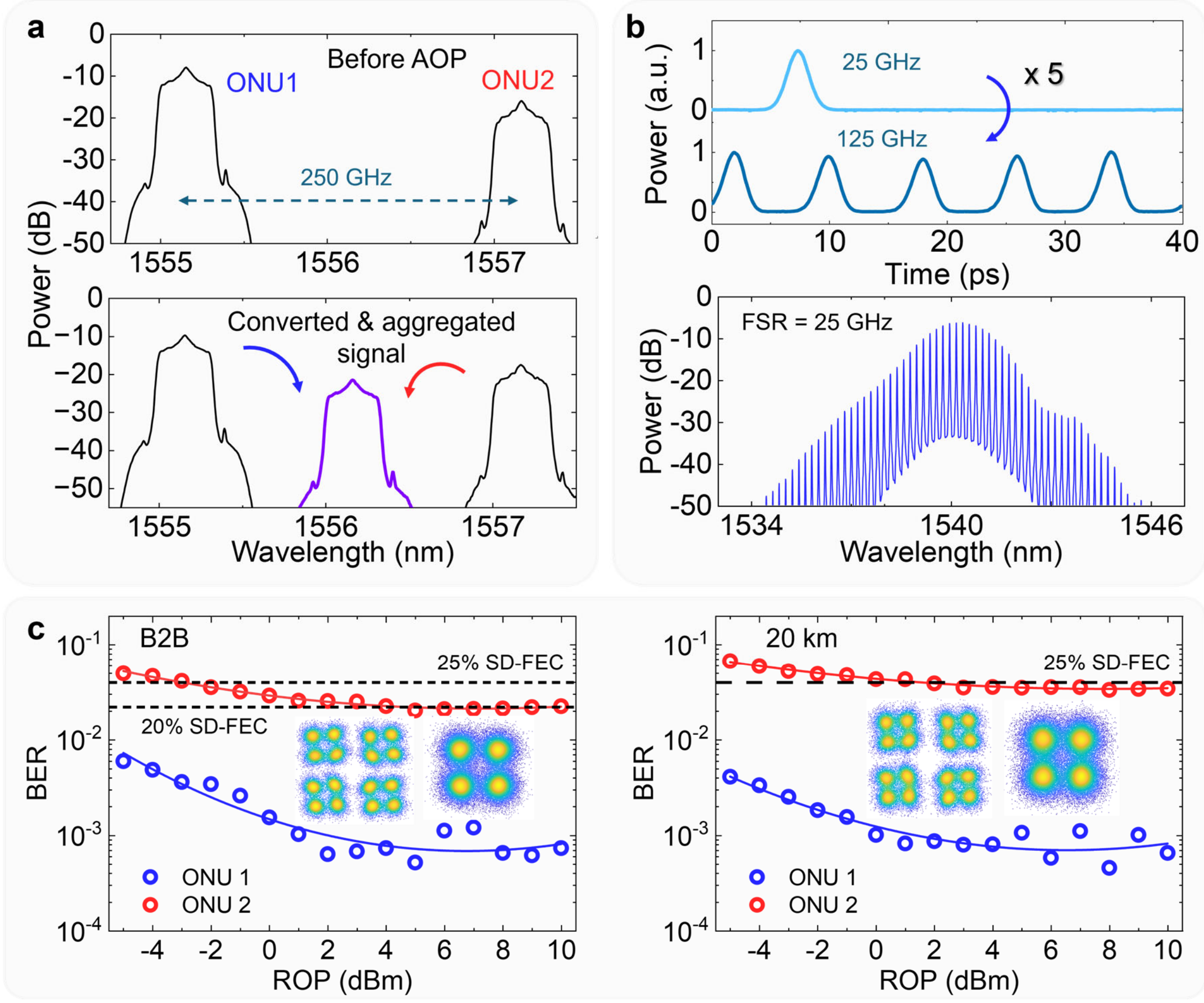

Figure 4 **a** Measured spectra before and after the all-optical processor (AOP). **b** Top: measured temporal traces before and after the temporal Talbot effect for $m = 5$. Bottom: Spectrum of the pulsed pump after the AOP. **c** Performance of ONU-1 and ONU-2 with B2B and 20-km transmission scenarios.

### 2.3. All-optical conversion of multiuser DSCM channels towards scalable edge uplink aggregations

Dense deployment of distributed edge nodes, including 6G open radio access network, radio units, edge computing servers, and campus access users, has driven explosive growth in MP2P upstream traffic, posing critical challenges for scalable, cost-effective, and spectrum-efficient coherent transmission [51]. Conventional WDM schemes rely on static wavelength allocation for each user group, while DSCM enables fine-grained multiuser access within a single wavelength, but both remain constrained by orthogonal multiplexing principles: spectral resources are statically reserved and cannot be flexibly reused, and central hubs require dedicated coherent receiver ports for each wavelength channel, leading to high capital and operational costs, excessive power consumption, and inflexible bandwidth scheduling. To address this challenge, we anchor our demonstration in a practical campus network scenario, where dense distributed users exchange frequent, high-capacity data and require reliable uplink connectivity to edge cloud resources, and experimentally validate our all-optical orthogonal-to-non-orthogonal multiplexing conversion scheme over a 3-km field-deployed campus fiber link.

To enable scalable multiuser uplink aggregation, we deploy our AOP at the edge aggregation node, where two WDM-separated DSCM channels, each carrying four fine-grained user subchannels (emulating 8 distributed edge users in total), are all-optically converted into a single power-superposed non-orthogonal DSCM channel prior to uplink transmission to the edge cloud (see Fig. 5a). Each DSCM channel employs four subcarriers, emulating realistic high-capacity edge access traffic. In conventional WDM-DSCM uplinks, the two wavelength channels would occupy exclusive spectral slots throughout the transmission link (even under low traffic load) and mandate two dedicated coherent receiver ports at the central hub. In contrast, our all-optical multiplexing format conversion releases these statically occupied wavelength slots, enabling spectral reuse for additional access nodes and reducing the number of required coherent receiver ports. This corresponds to a 2-to-1 consolidation in wavelength-slot usage and a 50% reduction in coherent receiver-port count at the aggregation node. This architecture preserves full compatibility with existing distributed user transmitters, eliminating the need for costly hardware upgrades at the user end and enabling smooth network evolution. While DSCM enables fine-grained bandwidth allocation for multiuser transmission, it still adheres to the orthogonal multiplexing principle, where spectral resources are statically reserved and cannot be flexibly reused across nodes.

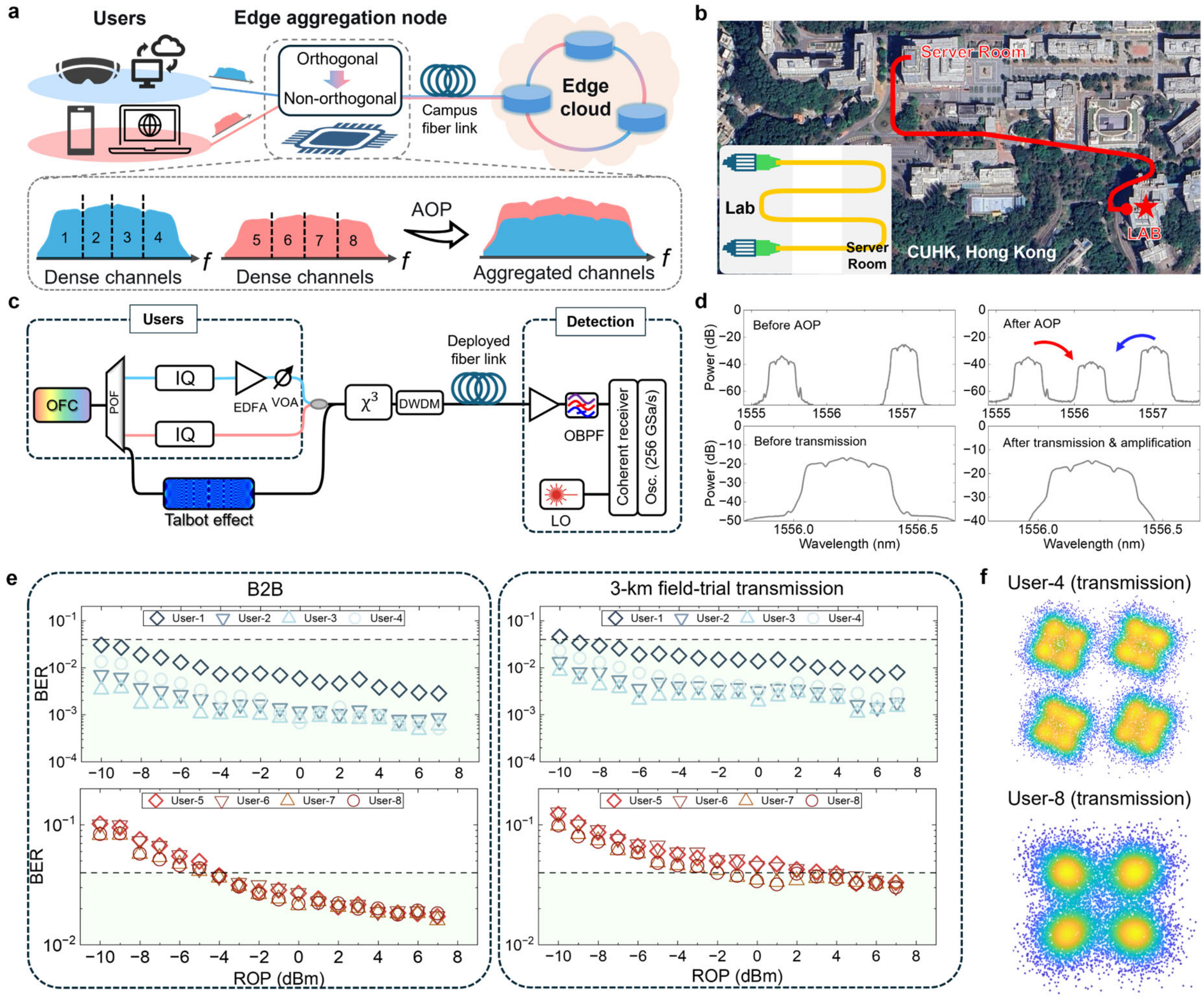

Figure 5 **a** Conceptual illustration of orthogonal-to-non-orthogonal multiplexing format conversion in DSCM-WDM upstream scenario. **b** Map of the deployed single mode fiber at the Chinese Univeristy of Hong Kong (CUHK). The 3-km field-deployed fiber link is realized by transmitting the signal twice between the laboratory and the campus server room. **c** Schematic illustration of experimental setup. OFC: optical frequency comb; POF: programmable optical filter; EDFA: erbium-doped fiber amplifier; VOA: variable optical attenuator; DWDM: dense wavelength-division multiplexer; OBPF: optical bandpass filter; LO: local oscillator. **d** Measured optical spectra showing the two input DSCM channels before the AOP, the converted DSCM-NOMA channel after the AOP, the DSCM-NOMA channel prior to uplink, and the DSCM-NOMA channel after field transmission and amplification at a received optical power of 7 dBm. **e** Bit error rate (BER) per subcarrier versus the received optical power (ROP) in both back-to-back (B2B) and fiber transmission scenarios. **f** Recovered constellation diagrams of User-4 (Top) and User-8 (bottom) after the field-deployed fiber transmission with an ROP of 7 dBm.

We experimentally validate this scheme over a 3-km field-deployed campus fiber link (Fig. 5b), which connects our laboratory to a campus server room (traversed twice) to emulate realistic uplink propagation conditions. Two optical carriers at 1555.35 nm and 1556.96 nm are modulated to generate the two 80 Gbit/s DSCM signals, with the 1556.96 nm carrier configured as the higher-power group and the 1555.35 nm carrier as the lower-power group. Subcarrier-pair Alamouti coding is adopted to enhance transmission robustness (see Supplement S6 and S7) [52]. The subcarriers on the higher-power carrier are labeled User-1 to User-4 (ordered from shorter to longer wavelengths), while those on the lower-power carrier are labeled User-5 to User-8. The launched powers into the nonlinear Kerr medium are approximately −16.6 dBm for the higher-power group and −23.7 dBm for the lower-power group. Again, an erbium-doped fiber amplifier (EDFA) followed by a VOA are used in the experiment to accurately set and control the power allocation ratio of the target aggregated channel.

The pulsed pump repetition rate is made programmable using a Talbot module composed of a programmable optical filter and a 3 km SMF. In our proof-of-concept experimental demonstration, the pulsed pump is derived from the same parametric comb source used to generate the optical carriers via spectral slicing. This configuration simplifies the experimental setup. In a practical deployment, however, the pump can be regenerated locally at the edge aggregation node where the AOP is placed [53, 54].

The multiplexing format conversion is performed prior to uplink transmission: before entering the deployed fiber link, the two DSCM signals are routed into the AOP, where Talbot-processed pulsed pumping and Kerr-mediated XPM coherently map the DSCM channels onto a single target center wavelength at 1556.16 nm. This operation produces a power-superposed, non-orthogonal DSCM channel, which is isolated by a DWDM filter and launched into the field-deployed campus fiber link. Figure 5d presents the measured optical spectra of the two input DSCM channels, the aggregated channel after AOP processing, and the aggregated channel after field transmission and amplification to a ROP of 7 dBm, respectively. The converted non-orthogonal signal is then captured by a single coherent receiver, and a tailored coherent DSP chain (augmented with SIC-based multiuser separation and Alamouti decoding) is applied to recover the information of all eight subcarriers (see Supplementary S8). System calibration identifies an optimal power allocation ratio of 7.5 dB for both B2B and field-deployed transmission cases (see Supplement S9). With this ratio fixed at 7.5 dB, the BER is measured as a function of ROP across all users. The BER of all users decreases monotonically with increasing received power, and for ROPs of 4 dBm or higher, all eight users achieve BERs below the 25% SD-FEC threshold. Representative constellation diagrams for User-4 (higher-power group) and User-8 (lower-power group) after field transmission at an ROP of 7 dBm are shown in Fig. 5f.

## 3. Discussion and conclusion

Orthogonal multiplexing over time and frequency has underpinned current high-capacity optical communication with rigorous and normative grids. Optical transport systems are now progressing toward ever higher baud rates per channel to satisfy emerging high-throughput applications. At the same time, the growing diversity and density of access services demand improved spectral utilization and network reconfigurability. Fixed orthogonal wavelength-grid WDM architectures therefore face increasing difficulty in scaling to next-generation optical networks, particularly in emerging scenarios that demand high-throughput optical signal processing. Introducing controlled non-orthogonality provides an alternative degree of freedom that can enhance flexibility and resource efficiency in future reconfigurable optical signal processing systems.

High-capacity orthogonal-to-non-orthogonal multiplexing format conversion requires all-optical signal processing capable of imposing ultrawideband, high-speed phase modulation, allowing XPM to mix widely separated carriers and high-baud signals [55, 56]. Our AOP meets this requirement by starting from a low-frequency RF-driven pulse train (25-GHz) and using programmable temporal Talbot processing to generate high-repetition-rate optical pulses (200-250 GHz). The Talbot processing, implemented with line-by-line spectral phase control and a dispersive medium, allows flexible repetition-rate multiplication ($m = 4$ and $m = 5$ in our demonstrations) and produces pump trains with repetition rates that can span a wide range. When these high-repetition-rate pumps co-propagate with target carriers through a Kerr nonlinear medium, the pump imposes ultra-wideband phase modulation on the channels via XPM. This mechanism enables coherent transfer of signals across large frequency gaps and supports processing of high-baud signals and dense multi-tone DSCM waveforms.

From an implementation perspective, the robustness of the proposed converter is supported by the comb architecture used in our proof-of-concept experiment, where both the pump and the signal carriers are derived from the same optical frequency comb and therefore share a common phase and timing reference [13]. Under this common-source configuration, the aggregated channels do not introduce additional relative frequency offsets after conversion, which significantly simplifies subsequent DSP-based multiuser recovery. However, when the pump and signal carriers are not mutually coherent, additional DSP procedures, such as pilot-assisted frequency offset estimation and phase noise compensation, are required to ensure reliable demodulation [57]. In addition, in our proof-of-concept demonstrations, the pump is operated in a moderate nonlinear regime, where the desired first-order converted component is sufficiently strong. The converted non-orthogonal multiplexing channel is subsequently selected by standard wavelength filtering after the nonlinear signal processing. The measured back-to-back and fiber-transmission BER results further confirm acceptable system performance within the operating window of our experiments.

The functional building blocks of the AOP are compact and amenable to integration. Specifically, a programmable optical filter, a dispersive element implementing Talbot mapping, and a $\chi^3$ waveguide is readily to be integrated into on-chip photonic platforms [58-61]. Integration would reduce insertion loss, improve thermal and phase stability, lower power consumption, and enable parallel replication of processors. Moreover, such integration enables a shared processing architecture, in which a single AOP module can serve multiple users at the aggregation node, thereby amortizing the hardware overhead across many channels. As a result, the overall system cost per user can be significantly reduced despite the added functionality of the optical-domain aggregation. From a system perspective, the converter can, in principle, be extended from two channels to a larger number of channels by mapping multiple orthogonal WDM carriers onto a single non-orthogonal power-domain channel. At the receiver side, the resulting multi-layer power-domain signal can then be recovered by sequential SIC, provided that the channels are arranged with a proper power hierarchy and sufficient separation between adjacent power levels. However, such multi-channel operation also introduces practical limitations, including stronger SIC dependence, stricter power-allocation requirements, and reduced tolerance to residual nonlinear products, spectral overlap, and signal-to-noise ratio degradation [62]. Therefore, practical scalability is achieved through modular and grouped aggregation, where multiple AOPs operate in parallel across the comb grid. In addition, when combined with an on-chip DWDM system [63], many AOPs can be tiled across the comb grid so that multiple orthogonal-to-non-orthogonal multiplexing format conversions can occur in parallel, providing a scalable substrate for future on-chip multi-channel photonic processors.

Beyond the proof-of-concept common-source configuration, the parametric OFC also provides a practical deployment model in which an OLT generates or regenerates a comb and distributes carriers to edge nodes [54, 64, 65]. When the pump and signal carriers are not mutually coherent, residual frequency offsets and phase drifts must be addressed [22]. Practical remedies include pilot tones and joint detection algorithms that explicitly account for frequency offsets [57]. Therefore, the absence of initial coherence between the WDM channels is not a fundamental barrier, but it moves complexity from the photonic aggregation stage into DSP, which can be implemented efficiently. The receiver-side synchronization and frame recovery are performed using standard coherent DSP [66].

In summary, our programmable all-optical orthogonal-to-non-orthogonal multiplexing format converter provides a practical approach to increase spectral efficiency and flexibility, with demonstrated applications in optical coherent networks. Our approach reclaims wavelength slots upstream of aggregation points, supports high-baud and widely spaced carriers, and interoperates with standard coherent receivers and DSP. Beyond optical communication networks, our technique can evolve from proof-of-concept demonstrations into a critical building block for next-generation metro and access networks, data-center interconnects, cloud-edge optical networks, and broader photonic information-processing platforms, enabling scalable, resource-efficient information aggregation in heterogeneous photonic systems.

## Acknowledgements

The authors thank CUHK Information Technology Services Centre for providing access to the campus fiber network for the experimental field trial. This work was supported by the Hong Kong RGC through GRF Grants 14221322, 14209423, 14211524, the NSFC/RGC Joint Research Scheme N_CUHK444/22, STG3/E-404/24-N, and RGC Junior Research Fellow Scheme (JRFS2526-5S09).

## Data Availability Statement

The data that support the findings of this study are available from the corresponding author upon reasonable request.